# Rényi Entropy Correction to Expanding Universe

H. R. Fazlollahi[1]

*Institute of Gravitation and Cosmology, Peoples Friendship University of Russia (RUDN University), 6 Miklukho-Maklaya St, Moscow, 117198, Russian Federation*

**Abstract**

The Rényi entropy comprises a group of data estimates that sums up the well-known Shannon entropy, acquiring a considerable lot of its properties. It appears as unqualified and restrictive entropy, relative entropy, or common data, and has found numerous applications in information theory. In the Rényi's argument, the area law of the black hole entropy plays a significant role. However, the total entropy can be modified by some quantum effects, motivated by the randomness of a system. In this note, by employing this modified entropy relation, we have derived corrections to Newton's law of gravitation. Taking this entropy associated with the apparent horizon of the Friedmann-Robertson-Walker Universe and assuming the first law of thermodynamics, $dE = T_A dS_A + WdV$, satisfies the apparent horizon, we have reconsidered expanding Universe and find modified Friedmann equations. Also, the second thermodynamics law has been examined.

According to Einstein's viewpoint, gravity is just the space-time curvature effects and is regarded as an emergent phenomenon that depicts the dynamics of space-time. After him, many attempts have been done to uncover the nature of the gravity field. One of the greatest steps in this direction was forwarded by Jacobson who by studying the thermodynamics of space-time and combining Clausius relation $\delta Q = T\delta S$ with the entropy expression illustrated explicitly that Einstein's field equation in general relativity is just an equation of state for the space-time [1]. It demonstrates that the Einstein field equations are the first law of thermodynamics for space-time. Following his investigations, a ton of studies have been done to reveal the profound association between gravity field and thermodynamics [2-4]. The considerations have been extended to the cosmological setups wherein it has been shown that the Friedmann equations in Friedmann-Robertson-Walker Universe can be written in the form of the first law of thermodynamics on the apparent horizon [5-9]. As discussed, in order to study and rewrite the Friedmann equations in any alternative or modified gravity theory through the first law of thermodynamics, $dE = T_A dS_A + WdV$, on the apparent horizon, we need to consider and apply the entropy expression of the black hole by replacing the black hole horizon radius $r_+$ by the apparent horizon radius $r_A$. In the past decades and with expanding Quantum mechanics concepts, the entropy expression associated with the black hole horizon are modified. These corrections on entropy expression include power-law and logarithmic corrections. In a general sense, the power-law corrections show up in dealing with the entanglement of quantum fields inside and outside the horizon [10-12]. A later form of corrections, namely the logarithmic corrections, arises from the loop quantum gravity due to thermal equilibrium fluctuations, quantum fluctuation, and uncertainty principle [13-15]. The Shannon entropy as an example of a logarithmic entropy class satisfies many operational quantities in information and communication theory in quantum mechanics [16]. However, in a non-asymptotic setting where the law of large numbers does not readily apply, it is broken and other entropy models such as collision entropy typically take over. Based on this, Rényi proposed new entropy expression nicely unifies these different and isolated measures in which entropy of a black hole area law can be modified as [17-19]

$$S_R = \frac{1}{\alpha}\ln(1 + \alpha S_0) \qquad (1)$$

where $S_0$ is the Bekenstein entropy, $\alpha$ is known as Rényi parameter. Clearly, for $\alpha \to 0$, the Bekenstein entropy is restored.

As the first step, it is worthwhile to reconsider Newton's law of gravitation and check how the extended form of entropy (1) modifies this law. In order to derive and modify Newton's law of gravitation, assume we have only two particles, one of them is the test particle with mass $m$, and the other as source one with mass $M$. To derive the entropic law, we define surface $S$ around source mass very close to the test particle as compared to its reduced Compton wavelength $\lambda_m = \hbar/(mc)$ and so the test particle is a distance $\Delta x = \eta \lambda_m$ away from the surface $S$. The entropy of the surface changes by the spectrum of the area of the surface via the relation ($\ell_p^2 = G\hbar/c^3$ is the Planck length),

$$\Delta S = \frac{\partial S}{\partial A}\Delta A = \left(\frac{1}{4\ell_p^2 + \alpha A}\right)\Delta A \qquad (2)$$

---
[1] seyed.hr.fazlollahi@gmail.com
Full name of author is seyed hamidreza fazlollahi.

The energy of the surface is identified with the relativistic rest mass of the source mass:

$$E = Mc^2 \qquad (3)$$

We also assume that surface $S$ is a set of bytes of information that scale proportional to the area of the surface,

$$A = \xi N, \qquad (4)$$

where $N$ represents the number of bytes and $\xi$ denotes the constant. Following the equipartition law of energy [20], the total energy on the surface as a function of temperature on the surface and bytes of information is, namely

$$E = \frac{1}{2} N k_B T. \qquad (5)$$

Supposing that the force on the test particle follows from the generic form of the entropic force governed by the thermodynamic equation of state such as

$$F = T \frac{\Delta S}{\Delta x}, \qquad (6)$$

As result, combining Eqs. (1)-(6) gives

$$F = \frac{Mm}{r^2} \left( \frac{\xi c^3}{\pi \eta k_B \hbar} \right) \left[ \frac{1}{4\ell_p^2 + \alpha A} \right]_{A=4\pi r^2} \qquad (7)$$

As shown, correction on entropy expression modified Newton's law of gravitation. Assuming $|\alpha| \ll 1$, Eq. (7) becomes

$$F \approx \frac{GMm}{r^2} \left( 1 - \frac{\alpha A}{4\ell_p^2} \right) = \frac{GMm}{r^2} - \frac{\alpha \pi GMm}{\ell_p^2}, \qquad (8)$$

where we set $\xi = 4\pi \eta k_B \hbar^2 / c^6$. Under this condition, the entropy-corrected Newton's law reduces to the usual Newton's law of gravitation only for $\alpha \to 0$.

In follows, we extend our discussion to the cosmological setup. We assume the background space-time is homogenous and isotropic described by the Friedmann-Robertson-Walker metric (we set $c = 1$)

$$ds^2 = h_{\mu\nu} dx^\mu dx^\nu + \tilde{r}^2 (d\theta^2 + \sin^2\theta d\phi^2), \qquad (9)$$

where $\tilde{r} = ra(t)$, $x^0 = t$, $x^1 = r$ and $h_{\mu\nu} = diag(-1, a^2/(1-kr^2))$ presents the two-dimensional metric. The $a(t)$ denotes the scale factor of the Universe and $k = 0, 1,$ and $-1$ correspond to the flat, closed and open Universe, respectively. The corresponding apparent horizon radius with FRW Universe, which is consistent with laws of thermodynamics, reads

$$\tilde{r}_A = \frac{1}{\sqrt{H^2 + k/a^2}} \qquad (10)$$

where $H = \dot{a}/a$ is the Hubble parameter of the Universe. Associated temperature with the apparent horizon can be defined as [21]

$$T_A = \frac{\kappa}{2\pi} = -\frac{1}{2\pi \tilde{r}_A} \left( 1 - \frac{\dot{\tilde{r}}_A}{2H\tilde{r}_A} \right) \qquad (11)$$

where $\kappa$ represents the surface gravity. the temperature on the apparent horizon becomes $T_A \leq 0$ for $\dot{\tilde{r}}_A \leq 2H\tilde{r}_A$. Thus to avoid the negative temperature one may consider $T_A = |\kappa|/2\pi$. The condition $\dot{\tilde{r}}_A \leq 2H\tilde{r}_A$ physically means that the apparent horizon is kept fixed and as result, there is no volume change. However, in this study, we consider Eq. (11), and not its absolute value and we will show that to keep the second thermodynamic law, one needs to have $\dot{\tilde{r}}_A > 2H\tilde{r}_A$. It demonstrates the profound connection between the temperature on the apparent horizon and the Hawking radiation [22].

As the simplest assumption, we consider the matter and energy content of the Universe as a form of a perfect fluid with stress-energy tensor

$$T_{\mu\nu} = (\rho + p) u_\mu u_\nu + p g_{\mu\nu} \qquad (12)$$

where $\rho$ and $p$ are the energy density and pressure, respectively. Using Bianchi's identity, $\nabla_\mu T^{\mu\nu} = 0$, leads to the continuity equation

$$\dot{\rho} + 3H(\rho + p) = 0 \qquad (13)$$

Following [23], the work density is work done due to the change in the volume, given by

$$W = -\frac{1}{2} T^{\mu\nu} h_{\mu\nu} \qquad (14)$$

Using Eq. (12) in Eq. (14) gives

$$W = \frac{1}{2}(\rho - p) \qquad (15)$$

Thus, the first law of thermodynamics on the apparent horizon recasts to

$$dE = T_A dS_A + W dV \qquad (16)$$

For a pure de Sitter space-time, $\rho = -p$, the work term reduces to the $-pdV$ and one reproduces the standard first law of thermodynamics $dE = T_A dS_A - pdV$.

If we suppose that the total energy content of the Universe inside a 3d sphere of radius $\tilde{r}_A$, given by $E = \rho V$, the differential form of total energy inside the apparent horizon becomes

$$dE = 4\pi \tilde{r}_A^2 \rho d\tilde{r}_A - 4\pi H \tilde{r}_A^3 (\rho + p) dt \qquad (17)$$

where Eq. (13) is used.

In order to study the effects of extended entropy forms, one only needs to consider extended entropy as the entropy associated with the apparent horizon. In this study, we consider Rényi entropy (1) as entropy on the apparent horizon.

$$dS_A = dS_R = \frac{2\pi \tilde{r}_A}{G + \alpha \pi \tilde{r}_A^2} d\tilde{r}_A. \quad (18)$$

As the final step in this setup, we substitute Eqs. (18), (15), and (11) in the first law of thermodynamics (16) and using relation (17),

$$4\pi H \tilde{r}_A^3 (\rho + p) dt = \frac{d\tilde{r}_A}{G + \alpha \pi \tilde{r}_A^2} \quad (19)$$

Integrating Eq. (19) gives:

$$\frac{4\pi}{3}\rho = \frac{1}{2G\tilde{r}_A^2} - \frac{\alpha\pi}{2G^2}(\ln(G + \alpha\pi \tilde{r}_A^2) - \ln(\tilde{r}_A^2)) \quad (20)$$

Substituting the definition of the apparent horizon radius (10), we get a modified version of the first Friedmann equation

$$H^2 + \frac{k}{a^2} - \frac{\alpha\pi}{G}\ln(\alpha\pi + G(H^2 + k/a^2)) = \frac{8\pi G}{3}\rho \quad (21)$$

In the limiting case where $\alpha = 0$, we recover the standard form of the first Friedmann equation.

As the first consideration, we would like to investigate the cosmological consequences of the modified Friedmann equation (21). Taking the time derivative of Eq. (21), yields

$$2H\left(\dot{H} - \frac{k}{a^2}\right) - 2\alpha\pi H\left(\frac{\dot{H} - k/a^2}{\alpha\pi + G(H^2 + k/a^2)}\right) = \frac{8\pi G}{3}\dot{\rho} \quad (22)$$

or by using continuity equation (13), we get

$$\dot{H} - \frac{k}{a^2} - \alpha\pi\left(\frac{\dot{H} - k/a^2}{\alpha\pi + G(H^2 + k/a^2)}\right) = -4\pi G(\rho + p) \quad (23)$$

After some calculations and using $\dot{H} = \ddot{a}/a - H^2$, one can rewrite the above equation as

$$2\frac{\ddot{a}}{a} + H^2 + \frac{k}{a^2} - \alpha\pi\left(\frac{3}{G}\ln(\alpha\pi + G(H^2 + k/a^2)) + 2\left(\frac{\dot{H} - k/a^2}{\alpha\pi + G(H^2 + k/a^2)}\right)\right) = -8\pi G p \quad (24)$$

This is the second modified Friedmann equation. For $\alpha = 0$, Eq. (24) shrinks to the usual second Friedmann equation,

$$2\frac{\ddot{a}}{a} + H^2 + \frac{k}{a^2} = -8\pi G p \quad (25)$$

Also, the second modified Friedmann equation can illustrate valuable ranges on Rényi parameter $\alpha$. Rearranging Eq. (24) with respect to $\ddot{a}/a$, gives:

$$\frac{\ddot{a}}{a} = -4\pi\rho\left(\frac{G}{3}(1 + 3\omega) + \frac{\alpha\pi(1+\omega)}{H^2 + k/a^2}\right) + \frac{\alpha\pi}{G}\ln(\alpha\pi + G(H^2 + k/a^2)) \quad (26)$$

where $\omega = p/\rho$ is the equation of the state of a perfect fluid. Since the current Universe is undergoing an acceleration phase, $\ddot{a} > 0$, from the above equation for $|\alpha| \ll 1$, we have three plausible sets of conditions: the prefect fluid behaves like the cosmological constant, $\omega = -1$. In this case, Rényi parameter is positive and smaller than the energy density, $\rho \gg \alpha$. This case represents the cosmological constant model with a small deviation due to constant $\alpha$. The other valuable condition arises from phantom-like fluid, $\omega < -1$, and $\rho > \alpha > -\frac{G(1+3\omega)(H^2+k/a^2)}{3\pi(1+\omega)}$. In this case, $\alpha$ take both positive and negative values. As the last possible case to keep an acceleration phase, we have quintessence-like matter in which $0 < \alpha < -\frac{G(1+3\omega)(H^2+k/a^2)}{3\pi(1+\omega)}$ and $-1 < \omega < -1/3$. It illustrates even by modifying Friedmann equations through the first law of thermodynamics and applying Rényi entropy as the effective entropy on the apparent horizon of the Universe, we have no restriction on the type of fluid and so one can explain accelerated expansion in the late-time era with different values on the equation of state $\omega$.

As the next step, it is worthwhile to consider the Universe wherein dominated fluid is the radiation, matter, and cosmological constant. For simplicity, we only consider the flat Universe, $k = 0$. However, one can extend the model to different values of $k$.

*Radiation-dominated era*:

Analyzing observations demonstrates the small random velocities of particles seen in the current Universe should have been a large part and governed the Universe in the past and some moments after the Big Bang [24]. As a result, the pressureless approximation for the early Universe breaks down. Due to high temperature, the Universe is filled with a highly relativistic gas, radiation, with the equation of state $p = \rho/3$. Under this condition, the continuity equation (13) becomes $\dot{\rho}_r + 4H\rho_r = 0$. Using Eq. (26) for this equation of state, we get

$$H\frac{dH}{dx} = -\frac{16\pi G}{3}\rho_r - \frac{16\pi G}{3}\rho_r(\alpha\pi + GH^2)^{-1} \quad (27)$$

where we use the *e-folding parameter* $x = \ln(a)$. This differential equation is a non-linear one and there is no exact solution. Hence, we approximate Eq. (27) due to the relation between the two parts of the last term. If $\alpha\pi \gg GH^2$, Eq. (27) becomes

$$H\frac{dH}{dx} \approx -\frac{16\pi G}{3}\rho_r(1 + G) + \frac{16G}{3\alpha}\rho_r H^2 \quad (28)$$

Solving it, yields

$$H^2 \approx c_1 + \alpha\pi + \frac{\alpha\pi}{G} - \frac{8G\rho_{r0}c_1}{3\alpha}e^{-4x} \quad (29)$$

where $c_1$ is the integration constant. Setting $c_1 = -\alpha\pi$, Eq. (29) recasts to

$$H^2 \approx \frac{8\pi G \rho_{r0}}{3} e^{-4x} + \frac{\alpha\pi}{G} \qquad (30)$$

which is the standard Friedmann equation in the radiation-dominated era with a deviation due to Rényi parameter. In another case, one can assume that $\alpha\pi \ll GH^2$ which gives,

$$H \frac{dH}{dx} \approx -\frac{16\pi G}{3} \rho_r \left(G + \frac{\alpha\pi}{GH^2}\right) \qquad (31)$$

For this case, the solution of Eq. (31) converges to Eq. (30), only for $\alpha < 0$, and becomes

$$H^2 \approx \frac{8G\rho_{r0}}{3} e^{-4x} - \frac{\alpha\pi}{G} \qquad (32)$$

In order to find the scale factor associated to expanding Universe, we solve Eqs. (30) or (32) using a perturbative method up to the only first order and considering a solution as

$$a = A_0 + \delta A_1 + \cdots \qquad (33)$$

where $A_0$ is the solution of (30) or (32) in absence of the second term on the right-hand side, $\alpha$ term, while the correction $\delta A_1$ satisfying the following equation,

$$H^2 = \pm \frac{\alpha\pi}{G} + \left(\frac{8G\rho_{r0}}{3} e^{-4x}\right)_{x=\ln(A_0)} \qquad (34)$$

The $A_0$ as the standard solution becomes,

$$A_0 = \tilde{c}_1 t^{1/2}, \qquad (35)$$

where $\tilde{c}_1 = (32\pi G \rho_{r0}/3)^{1/4}$. Substituting the above relation into Eq. (34), and using relation (33), we get

$$a \approx \tilde{c}_1 t^{1/2} + c_2 \chi \left(1 - \frac{\pi \tilde{c}_1^2 t^2}{G\sqrt{\tilde{c}_1}} \alpha\right) \qquad (36)$$

where $\chi_1 = \left(\frac{4\tilde{c}_1 G^2}{et}\right)^{\frac{1}{\tilde{c}_1\sqrt{\tilde{c}_1}}}$ and $c_2$ is the integration constant. The first term in Eq. (36) is the standard scale factor form for the radiation-dominated era while the second term arises from the correction of Bekenstein entropy, Eq. (1). While the first term has a key role in the evolution of scale factor for $t > 1$, the second term demonstrates radiation-dominated era could not be the first era after Big Bang and plays the main role for $t < 1$. As result, Rényi corrections on entropy lead one to inflation era in very early Universe, before radiation-dominated era.

*Matter-dominated era*:

It is notable that the pressure caused by randomness motions of cosmic systems, galaxies, and clusters, is negligible and is of the order of $10^8$cm s$^{-1}$ or less [24]. This implies that the pressure of matter, baryonic and dark matter, is negligible compared to relativistic fluid. Thus, if the Universe is filled only by non-relativistic matter with negligible pressure, we can assume $p_m = 0$, and so the continuity equation recasts to $\dot{\rho}_m + 3H\rho_m = 0$. For such fluid, pressureless matter, Eq. (26) becomes

$$\frac{\ddot{a}}{a} = -4\pi G \rho_m \left(1 + \frac{\alpha\pi}{\alpha\pi + GH^2}\right) + H^2 \qquad (37)$$

Rewriting the above equation by using $\frac{\ddot{a}}{a} = \dot{H} + H^2$ and *e-folding parameter* $x = \ln(a)$, gives

$$H \frac{dH}{dx} = -4\pi G \rho_m - 4\alpha G \pi^2 \rho_m (\alpha\pi + GH^2)^{-1} \qquad (38)$$

By applying the same approach that was used for the radiation-dominated era, one gets

$$H^2 \approx \frac{8\pi G \rho_{m0}}{3} e^{-3x} \pm \frac{\alpha\pi}{G} \qquad (39)$$

where plus and minus signs are corresponding to positive and negative $\alpha$, respectively. using the perturbative method (33) gives the scale factor

$$a \approx \tilde{c}_2 t^{2/3} - c_3 \chi_2 \left(1 - \frac{\pi \tilde{c}_2^3 t^{8/3}}{14G} \alpha\right) \qquad (40)$$

where $\tilde{c}_2 = (6\pi G \rho_{m0})^{1/3}$, $\chi_2 = 3\left(\tilde{c}_2^{9/2} t\right)^{-1/3}$ and $c_3$ is an integration constant of the model. The coefficient $\chi_2$ vanished for large cosmic time and so the second term in Eq. (40) makes no tangible effects at end of the matter-dominated era. However, this term can play important role in the first steps of the transition from radiation to a matter-dominated era.

Computing Hubble parameter for $|\alpha| \ll 1$, and rearranging it gives,

$$\frac{1}{H} \approx \frac{3t}{2} - \frac{9c_3}{2\tilde{c}_2^{5/2}} + \frac{3\alpha\pi c_3 \chi_2 \tilde{c}_2^2}{28G} t^3 \qquad (41)$$

The two last terms arise from entropy modification. To investigate the effects of these terms, we can estimate Rényi parameter at current Universe, $t = t_0$, as

$$\frac{1}{H_0} \approx \frac{3t_0}{2} - \frac{9c_3}{2\tilde{c}_2^{5/2}} + \frac{3\alpha\pi c_3 \chi_2 \tilde{c}_2^2}{28G} t_0^3, \qquad (42)$$

where $H_0 = H(t_0)$ is the Hubble constant. Ignoring extra terms, two last terms give the $t_0 = 2/(3H_0)$ which is the age of the Universe in standard cosmology [25]. Using definition of $\chi_2$, and ignoring the second term for simplicity, Eq. (42) becomes

$$\frac{1}{H_0} \approx \frac{3t_0}{2}\left(1 + \frac{3\alpha\pi c_3\sqrt{\tilde{c}_2}}{14G}t_0^{5/3}\right) \tag{43}$$

The Hubble constant is usually given as a function of uncertainty $h$ such as

$$H_0 = 100h \text{ km s}^{-1} \text{ Mpc} \approx 2.1332h \times 10^{-42} \text{ GeV} \tag{44}$$

The observations evaluate $h = 0.72 \pm 0.08$ [26] and thus the age of the Universe becomes $t_0 \approx 13.4 \, Gy$. Substituting these parameters into Eq. (43) shows $\alpha \sim 10^{-29 \pm 1}$ when we set $c_3\sqrt{\tilde{c}_2}/G = 1$. As result, it shows Rényi parameter $\alpha$ has no effective role in the age of the Universe. However, it plays an outstanding role in the beginning steps of radiation and the transition point between radiation and the matter-dominated era.

*Cosmological Constant model*:

Although the cosmological constant model suffers from some problems [27], it is considered the simplest theory to describe the dark energy era, a late-time epoch, of the Universe. In this model, the equation of state $p = -\rho$ and thus the energy density does not evolve with time and is $\rho \approx 3.34 \text{ GeV m}^{-3}$ [28]. For such fluid, Eq. (26) becomes

$$\frac{\ddot{a}}{a} = H^2, \tag{45}$$

Solving this differential equation gives:

$$a = a_0 e^{c_3 t} \tag{46}$$

It implies filling Universe only with cosmological constant yields inflation expansion phase. This is the straightforward result when one studies Universe in which the cosmological constant is considered a dominated fluid [29].

Using the first or second Friedmann equation gives,

$$\rho = \frac{3c_3^2}{8\pi G} - \frac{3\alpha}{8G^2}\ln(\alpha\pi + Gc_3^2) = -p \tag{47}$$

which shows may Rényi parameter $\alpha$ can solve some inconsistencies in the value of energy density of cosmological constant [27].

Till now, with Rényi entropy (1) and the first law of thermodynamics, we derive the first and second Friedmann equations. Also, the evolution of the Universe is considered under some specific values of the equation of state. In the last part of this study, we examine the second law of thermodynamics. Rewriting Eq. (19), we find

$$4\pi H\tilde{r}_A^3(\rho + p) = \frac{\dot{\tilde{r}}_A}{G + \alpha\pi\tilde{r}_A^2} \tag{48}$$

Solving the above equation for $\dot{\tilde{r}}_A$, gets

$$\dot{\tilde{r}}_A = 4\pi H\tilde{r}_A^3(G + \alpha\pi\tilde{r}_A^2)(\rho + p) \tag{49}$$

The sign of $\dot{\tilde{r}}_A$ depends on the equation of state of the fluid, only when $\alpha \ll G$. When the energy condition holds, $\rho + p > 0$, we have $\dot{\tilde{r}}_A > 0$. To explore the second law of thermodynamics, $T_A\dot{S}_A$ becomes,

$$T_A\dot{S}_A = -\frac{1}{2\alpha\pi\tilde{r}_A}\left(1 - \frac{\dot{\tilde{r}}_A}{2H\tilde{r}_A}\right)\frac{d}{dt}(\ln(1 + \alpha S_0)) \tag{50}$$

Taking time derivative and using Eq. (49) into Eq. (50), gives

$$T_A\dot{S}_A = -4\pi H\rho\tilde{r}_A^3\left(1 - \frac{\dot{\tilde{r}}_A}{2H\tilde{r}_A}\right)(1 + \omega) \tag{51}$$

It implies the second law of thermodynamics is satisfied only under two sets of conditions $\omega < -1$ and $\dot{\tilde{r}}_A < 2H\tilde{r}_A$ or when $\omega > -1$ while $\dot{\tilde{r}}_A > 2H\tilde{r}_A$. The first set of conditions presents a phantom-like fluid and the second one denotes fluid as a quintessence field. Under other sets of conditions the second law of thermodynamics, $\dot{S}_A \geq 0$, does not hold on the apparent horizon of the system, the Universe. To better insight, we need to examine the validity of the generalized second law of thermodynamics, namely $\dot{S}_A + \dot{S}_m \geq 0$.

From the Gibbs equation, we have [30]

$$T_m dS_m = dE + pdV = Vd\rho + (\rho + p)dV \tag{52}$$

where $T_m$ and $S_m$ are temperature and entropy of the matter fields inside the apparent horizon of the Universe, respectively. To continue our discussion, we assume that the local equilibrium hypothesis holds. As result, the temperature inside of the apparent horizon of the Universe remains in equilibrium and thus the temperature of the Universe must be uniform and the same as the temperature of its boundary, the apparent horizon, which implies $T_m = T_A$ [30]. Thus, the Gibbs equation (52) becomes

$$T_A\dot{S}_m = 4\pi\tilde{r}_A^2\dot{\tilde{r}}_A(\rho + p) - 4\pi\tilde{r}_A^3 H(\rho + p) \tag{53}$$

In order to examine the generalized second law of thermodynamics, we must study the evolution of the total entropy $\dot{S}_A + \dot{S}_m$. Adding Eq. (51) to Eq. (53), yields

$$T_A(\dot{S}_A + \dot{S}_m) = 2\pi\rho\tilde{r}_A^2(1 + \omega)(3\dot{\tilde{r}}_A - 4\tilde{r}_A H) \tag{54}$$

To fulfill the second law of thermodynamics, Eq. (54) must be a non-decreasing function of time. Therefore, for quintessence-like fluid, $\omega > -1$, we should have $\dot{\tilde{r}}_A > 4\tilde{r}_A H/3$. Only under this condition, the temperature on the boundary, apparent horizon, be positive while for the phantom-like field, $\omega < -1$, the second law of thermodynamics holds for $\dot{\tilde{r}}_A < 4\tilde{r}_A H/3$. This case gives negative temperature on boundaries and so is not a valuable one. It demonstrates to keep the validity of the generalized second law of thermodynamics when $T \geq 0$, only quintessence-like fields, $\omega \geq -1$, are expected.

In the conclusion, we have reconsidered the expanded Universe and Newton's law of gravitation due to the correction of the expression of entropy. Although the standard entropy is modified due to different aspects of quantum properties, in this study Rényi entropy is used which arises from quantum information theory. As shown, the classical gravitation equation, Newton's law of gravitation, modified depending on Rényi parameter. Investigating Friedmann equations shows they are modified due to entropy expression. Examining the age of the Universe suggests that Rényi parameter is in order of Planck constant and plays a key role in the first steps of evolution of radiation-dominated era and transition point from radiation to matter-dominated era. Exploring the Universe filled with cosmological constant, $\omega = -1$, gives exponential scale factor while energy density includes two terms, the first term depends on scale factor while the second term arises from entropy correction. This term may alleviate some inconsistencies between theoretical and observational values of vacuum energy. Also, we have examined the total generalized second law of thermodynamics which implies that this law holds only for quintessence fluids, $\omega > -1$.

## Acknowledgment

The author thanks V. D. Ivashchuk and A. H. Fazlollahi for their helpful comments and review.